\documentclass[10pt]{article}
\usepackage[pass]{geometry}
\usepackage{times}
\pdfoutput=1

\usepackage{epsfig}

\begin{document}

\baselineskip=12 pt

\title{GRB: Greedy Routing Protocol with Backtracking for Mobile Ad
  Hoc Networks\\ (Extended Version)}


\author{Baban A. Mahmood \\University of Kirkuk\\
Kirkuk,  Iraq. \\Email: Baban@netlab.uky.edu \and D. Manivannan
   \\ University of Kentucky\\ Lexington, Kentucky, USA.\\ Email: Mani@cs.uky.edu}

\maketitle

\begin{abstract}

Routing protocols for Mobile Ad Hoc Networks (MANETs) have been
extensively studied for more than fifteen years. Position-based
routing protocols route packets towards the destination using greedy forwarding (i.e., an intermediate
node forwards packets to a neighbor that is closer to the destination
than itself). Different position-based protocols use different strategies to pick the
neighbor to forward the packet. If a node has no neighbor
that  is closer to the destination than itself, greedy forwarding
fails. In this case, we say there is void (no neighboring nodes) in
the direction of the destination. Different position-based routing protocols
use different methods for dealing with voids. In this paper, we use
a simple backtracking technique to deal with voids and design a position-based
routing protocol called ``Greedy Routing Protocol with Backtracking (GRB)''. We
compare the performance of our protocol with the well known  Greedy
Perimeter Stateless Routing (GPSR) routing and the Ad-Hoc
On-demand Distance Vector (AODV) routing protocol as well as the Dynamic Source Routing (DSR) protocol. Our protocol needs
much less routing-control packets than those needed by DSR, AODV, and
GPSR. Simulation results also show that our protocol has a higher
packet-delivery ratio, lower  end-to-end delay, and less hop count on
average  than  AODV.

\end{abstract}



\section{Introduction}\label{Introduction}

A Mobile Ad-hoc Network (MANET) consists of a set of nodes each of
which is capable of being both a host and a router.  The nodes form  a
network among themselves  without the  use of any fixed
infrastructure, and communicate with each other  by  cooperatively
forwarding packets on behalf of others.  Mobile ad-hoc  networks have
applications in areas such as  military, disaster  rescue operations,
monitoring animal habitats, etc. where  establishing communication
infrastructure is not  feasible~\cite{HuaizhiLi_Baban_2005,HShen_Baban_2013, NazariTV_Baban_2013,KKavithal_Baban_2013,JainSS_Baban_2013}. Routing  protocols designed
for  mobile ad hoc networks  need to be   scalable, robust, and have
low routing overhead. Routing protocols   designed for MANETs can be
broadly classified as geographic routing   protocols (or
position-based routing protocols) and
topology-based routing   protocols. In geographic   routing protocols,
nodes do not maintain information related to   network topology (i.e.,
they are topology independent). They only depend   on the location
information of nodes to make forwarding
decisions. Generally~\cite{junlongLin_Baban_2006}, nodes need their own
location, their neighbors' location, and the location of the
destination node to which the packet needs to be forwarded. Using
this location information, routing is accomplished by forwarding
packets hop-by-hop until the destination node is
reached~\cite{RolandF_Baban_2008}. Greedy forwarding (GPSR~\cite{BradKarp_Baban_2000}), is one
of the main strategies used in geographic   routing protocols. Under
Greedy forwarding, an intermediate node on the   route forwards
packets  to the next neighbor  node that is closer to   the
destination than   itself.

Topology-based routing protocols depend on current topology of
 the  network. Topology-based routing  is also known as table-based
 routing. Topology-based routing can be  classified in to proactive
 routing protocols, reactive routing  (on-demand) protocols, and
 hybrid routing protocols~\cite{HuaizhiLi_Baban_2005,ChiaCH_Baban_2009,VMukesh_Baban_2007}.
In proactive protocols, like DSDV~\cite{CalresEP_Baban_1994}, nodes use
pre-established table-based routes~\cite{StefanoBI_Baban_1998}. Therefore, routes
are deemed reliable and nodes do not wait for route discovery which
cuts off latency. However, overhead incurred for route
construction and maintenance can degrade performance, limit
scalability, and the routing table will consume lot of memory  as the
network size grows.

   Reactive Routing Protocols are also called on-demand routing
 protocols  wherein  senders find and maintain  route to a destination
 only when  they need it. Reactive routing needs less memory and
 storage capacity than  proactive protocols. However, in network areas
 where nodes can move  more unpredictably and frequently, path
 discovery may fail since the  path can be long and links may break
 due to node mobility or when  facing other obstacles~\cite {HuaizhiLi_Baban_2005}. The delay caused by  route discovery for each data
 traffic can increase latency. 

On the other hand, geographic routing protocols require only the
location information of  nodes for routing. They do not require a node
to establish a route to the  destination before transmitting
packets. Unlike on-demand routing  protocols, they do not depend on
flooding route request messages to discover routes. This feature
helps geographic routing protocols to reduce the extra overhead
imposed by topology  constraints for route
discovery~\cite{zhaoyb_Baban_2007,RolandF_Baban_2008}.  A node only needs to know the
position of its neighbors and the position of the destination to
forward packets. Therefore, geographic routing protocols generally are
more  scalable than topology  based routing
protocols~\cite{lemmoncls_Baban_2009,shobanaks_Baban2013,cadgerfc_Baban_2013}. 
In spite of the benefits mentioned above,  geographic routing
protocols have the following  limitations: Greedy forwarding, the
primary packet forwarding strategy  used by  geographic routing
protocols,  may fail in low density networks,  networks with
non-uniformly distributed nodes, and/or networks where obstacles can
be present. Moreover, Location  Service is required to obtain
location information  of destination  nodes which may result  in
high  overhead. The non-hierarchical address structure used in
ad-hoc networks requires more  control  overhead to update  node
location~\cite{lemmoncls_Baban_2009}. 

\subsection*{Paper Objective}
Many geographic routing protocols construct the planarized version of
the local network graph to route a packet around voids; constructing
planarized graph of the local network requires exchanging neighborhood
information of nodes at least two hops away and then planarizing the
graph,  which can cause large overhead, especially in  a sparse
network wherein several voids may exist on  a route. In addition to that, planarization may fail to generate bidirectional, connected, and/or cross link free local graphs as observed by Kim et al. and Frey et al.~\cite{Kimyrkb_Baban_2005,freyFace_Baban_2006}. In this paper, we
address this issue and propose a simple geographic routing protocol
that uses backtracking to route packets around voids.
  
\subsection*{Organization of the Paper}
The rest of the paper is organized as follows. In
Sect.~\ref{RelatedWork}, we discuss the related work and paper objectives. In
Sect.~\ref{algorithm}, we present  our protocol. In Sect.~\ref{Analysis}, we present the performance
evaluation results of our protocol. In Sect.~\ref{Discussion} we give a brief discussion of our protocol. Sect.~\ref{Conclusion} concludes the paper.

\begin{figure*}[t]
	 \centering
	 \includegraphics[width=0.6\textwidth]{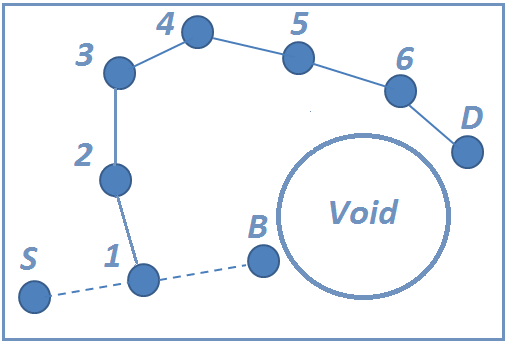}
	 \caption{ Example Illustrating Dead End (void) in Greedy Forwarding.}
	  \label{fig:DeadEnd}
 \end{figure*}

\section{Related Work}\label{RelatedWork}  
In this section, we discuss the basic idea behind several  of the
recently proposed geographic (position-based)  routing protocols  and on demand routing protocols bringing out their
week and strong points.

\subsection{Position-based Routing Protocols}
GPSR~\cite{BradKarp_Baban_2000}, a well known geographic routing
protocol proposed by  Karp and Kung, uses greedy forwarding as the
default forwarding strategy. When a packet confronts a void (i.e.,
when greedy forwarding fails), they planarize the local topology graph
either by constructing the Relative Neighborhood Graph (RNG) or
Gabriel Graph (GG) of the local network graph and use those graphs to
route around the void.   However, constructing such graphs involves overhead.  

Zhao et al.~\cite {zhaoyb_Baban_2007}, proposed a routing protocol
called HIR. HIR selects specific nodes as landmarks and builds a
multidimensional coordinate system based on which they find the Hop ID
distance between each pair of nodes in the network. When a packet
meets with a void, the protocol switches to a landmark-guided, detour
routing  which tries to forward the packet to the landmark nearest to
the  destination. This protocol does not scale  well due to the
central packet flooding technique used by a  designated node to select
LANDMARKs. Nodes  proportionally exchange high amount of control
information to select LANDMARKs. However, since this happens only to
select LANDMARKs, normal data packet forwarding process does not
produce additional overhead. On the other hand, the level of stability
of LANDMARK nodes plays a significant role in control   overhead.  

Li and Singhal~\cite {HuaizhiLi_Baban_2005}  proposed ARPC in which the routing process is divided into several parts. First, Location-based Clustering Protocol in which several physical locations  are assumed to be known in the network area which are called anchors and they have coordinates. Second, Inter-cell Routing Protocol which lets every node maintain a dynamic routing table that contains routes to its neighboring cells. Third, Intra-cell Routing Protocol which is an on-demand routing technique that is performed inside the same cell. Fourth, Data Packet Routing which deals with  how the data packets are routed  from a source to a destination node. ARPC is less scalable since  nodes use routing tables which can become large  as the network size increases. Moreover,  the announcement packets sent by agent nodes  to announce their existence to the nodes in the cell they reside  can cause large overhead as the node density increases. 

Lin and Kus proposed LGR~\cite{junlongLin_Baban_2006}  which is a location-fault-tolerant  geographic routing protocol that uses both traditional geographic  routing and position-based clustering technique. Routing is performed using global geographic routing and local  gradient routing. A cluster head  (CH) broadcasts  messages to all the nodes in its own cluster as well  as all nodes in its neighboring clusters so that each of these nodes can establish a routing  path to the CH. This process affects scalability. If nodes  are  highly mobile, frequent CH election occurs which results in more  message broadcasting by newly elected CHs to announce their existence which overwhelms nodes with control packets exchanged between nodes and hence limits scalability. 

Zhou et al. presented Geo-DFR~\cite {zhoublyzg_Baban_2008} which
incorporates  directional forwarding in routing
(DFR)~\cite{gerlamlyzz_Baban_2006}. Routing is done mainly using
greedy forwarding. However, in case of dead ends (voids), DFR  is
used. Geo-DFR improves DFR to solve the dead end  problem and avoids face routing. The authors use Fisheye State Routing protocol (FSR)~\cite{GuangyuGr_Baban_2000} which is the protocol that is "hosting" Geo-DFR. Scalability of Geo-DFR is affected by different factors. First, maintaining three tables in  each node increases overheard especially when a node has many  neighbors. But this limitation is local, since the number of records in two of these tables depends on the density of the neighboring nodes. 

Li et al.~\cite{XLINMN_Baban_2012} proposed localized load-aware geographic routing using the concept of cost-to-progress ratio in greedy routing  (CPR-Routing). The main idea behind this protocol  is to combine the greedy forwarding technique and localized cost-to-progress ratio (CPR)~\cite{StojmenovicI_Baban_2006}. The load awareness used in this protocol tends to minimize load and maximize progress geographically towards the destination; however,  this is difficult to achieve, so it tries to balance the two factors  in making routing decisions. A drawback of this approach is  the complexity involved in the calculation for selecting appropriate neighbor to forward a packet which could result in higher end-to-end delay. 

Macintosh  et al. proposed LANDY~\cite{macintoshsfl_Baban2012} which uses locomotion (movement) and velocity of each node to predict  the future location of each of these nodes so that data packets are  forwarded efficiently towards the destination nodes. LANDY uses  only local broadcasting to build a Locomotion Table (LT). When forwarding fails,  instead of broadcasting, a recovery mode is invoked from the point of failure, allowing the protocol scale better. Overhead involved in  this protocol is higher than that of the  protocols discussed so far since it uses more control information to build tables at each node. That includes different samples of each node's location information exchanged periodically between nodes and building planar graphs to be used as alternatives to the normal forwarding mode.

\subsection{On-demand Routing Protocols}

We compare our  protocol with two of the well-known on-demand routing protocols that are discussed bellow. Under AODV~\cite{CalesEPS_Baban_1999}, when a node needs to establish a route to a
destination, it  broadcasts a route request to all its neighbor
nodes. A node receiving the route request node replies to the source
node, if it has a route to the destination; otherwise, it rebroadcasts
the route request to all its neighbors. This process continues until
the a route to the destination is found. This protocol is robust
because broadcasting route request guarantees finding a route to the
destination if there is one; however, as number of nodes increase, the
number of redundant rebroadcasting of route requests increases. This
means this protocol  is not  scalable. 

Under DSR~\cite{johnsondm_Baban_1996}, when a node $S$ needs to find a
route to a node D, it broadcasts a Route Request packet (RREQ). On
receiving RREQ, an intermediate node adds its id  to the list of nodes
in RREQ if it has no route to the destination $D$ and if its id is
not in already in the list; then it rebroadcasts the updated RREQ
packet. When a target node receives the  RREQ, it puts the
 list of nodes received in RREQ   in the Route Reply (RREP)  and sends
 it back to the sender S. When $S$ receives the RREP, the sender caches
 the route for subsequent routing. This protocol also results in
 redundant propagation of route requests.

\section{Problem Statement and Design of our Protocol}\label{algorithm}

In this section, we first present the objective of the paper and the
basic idea behind our protocol, and then present a detailed description of the protocol.

\subsection{Problems Addressed and Solved in This Paper}\label{problemstatement}

Greedy forwarding, the primary packet forwarding strategy used by geographic routing
protocols, may fail in low density networks, networks with non-uniformly distributed nodes, and/or networks where obstacles can be present. Therefore, the main problem with greedy forwarding strategy is that it does not guarantee packet delivery to the destination because of the dead end phenomenon even if there is a route to the destination. Figure~\ref{fig:DeadEnd} shows an example of dead end (void) problem. When the source node $S$ needs to send packets to the destination node $D$, it
forwards the packets greedily to a node that is closer to the destination than itself. On
receiving the packets, each subsequent node does the same. When the packets reach node
$B$, it finds that none of its neighbors are closer to the destination node $D$ than itself and $D$ is
outside the transmission range of node $B$. This means there is a void between $B$ and $D$ as depicted in Figure~\ref{fig:DeadEnd}.
Even though there is a valid path from $S$ to $D$ through intermediate nodes \emph{1,2,3,4,5,} and \emph{6}, greedy forwarding cannot use it. This means, under pure greedy forwarding, packets may be dropped even though there are valid
paths to destination nodes. 

Generally, position-based routing protocols use planarization and face routing~\cite{BradKarp_Baban_2000,lemmoncls_Baban_2009} to go around voids. Planarization involves constructing the planar graph of local network. Graphs are generally planarized using the Gabreil Graph (GG)~\cite{gabrielks_Baban1969} or the Relative Neighborhood Graph (RNG)~\cite{toussaintGo_Baban_1980}. However, planarization may fail to generate bidirectional, connected, and/or cross link free local graphs as observed by Kim et al. and Frey et al.~\cite{Kimyrkb_Baban_2005,freyFace_Baban_2006}. This may be the result of node's incorrect estimate of its location or irregular communication range as a result of radio-opaque obstacles or transceiver differences.

 \begin{figure}[h]
	 \centering
	 \includegraphics[width=0.5\textwidth]{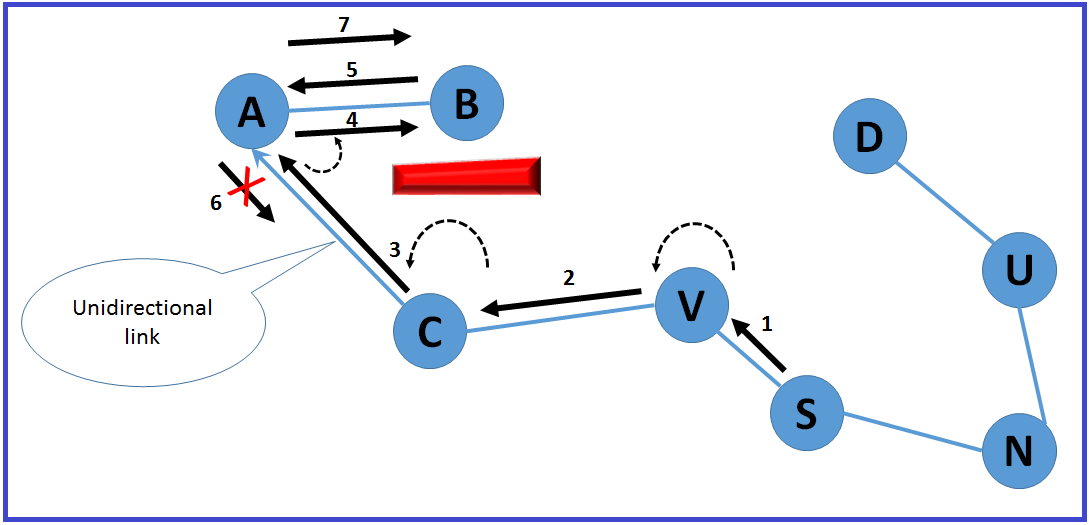}
	 \caption{ Example of Planarization where a Unidirectional Link Causes Routing Failure.}
	  \label{fig:PlanarizatoinUnidirectional}
 \end{figure}

As shown in Figure~\ref{fig:PlanarizatoinUnidirectional}, a unidirectional link can cause an infinite loop during face traversal. In this example, when the source node $S$ needs to send data packets to the destination node $D$, based on greedy forwarding, it sends that data packet to node $V$. Node $V$ is a dead end for that packet, hence it switches to face routing and forwards the data packets to node $C$. In this case, when $C$ constructs the planar graph of the local network using $GG$, it cannot see the witness $B$ in the circle whose diameter is the distance between $A$ and $C$ because of the obstacle shown in Figure~\ref{fig:PlanarizatoinUnidirectional}. Therefore, $C$ generates a link to node $A$. However, node $A$ does not create a link to node $C$ in its local graph because it can see the witness $B$ in the circle. Therefore, node $C$ can forward the data packets to node $A$ which in turn forwards them to node $B$ and based on face routing, node $B$ returns the packet to node $A$. Since node $A$ does not have a link to node $C$ in the local graph, it returns the packets to node $B$ and as a result, the data packets loop.

 \begin{figure}[h]
	 \centering
	 \includegraphics[width=0.65\textwidth]{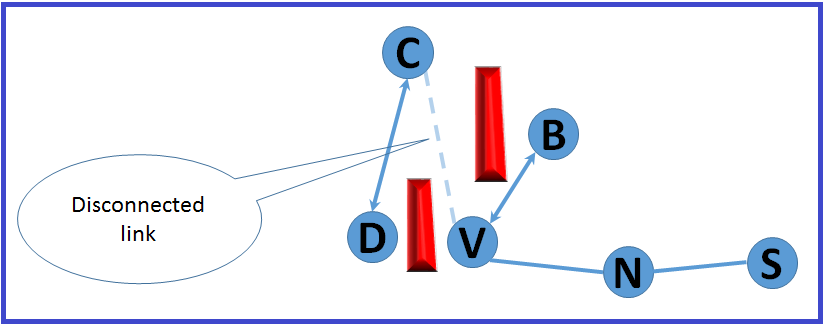}
	 \caption{ Example of Planarization where a Disconnected Link Causes Routing Failure.}
	  \label{fig:PlanarizatoinDisconnected}
 \end{figure}

Routing can also fail because of disconnected links as shown in Figure~\ref{fig:PlanarizatoinDisconnected}. In this example, source $S$ needs to send data packets to destination $D$. When packets arrive at node $V$, they face a dead end and switch to face routing. From node $V$'s view, $B$ is a witness and from $C$'s view, $D$ is a witness. Therefore the link between $V$ and $C$ is removed in the planarized graph of local network and as a result, the local graph is disconnected and packets cannot travel to $D$.

 \begin{figure}[h]
	 \centering
	 \includegraphics[width=0.73\textwidth]{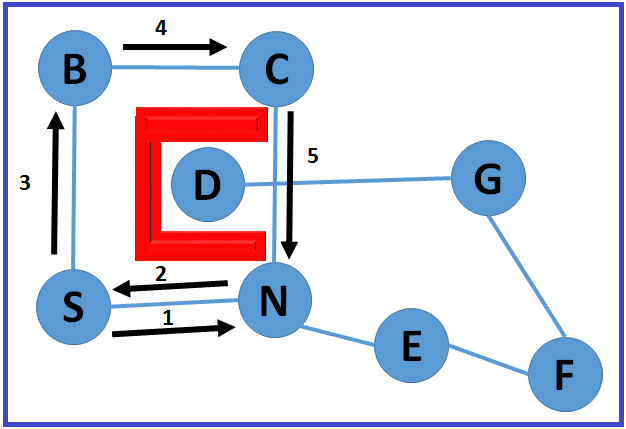}
	 \caption{ Example of Planarization where a Cross Link Causes Routing Failure.}
	  \label{fig:PlanarizatoinCrossLink}
 \end{figure}

Another scenario wherein routing may fail is shown in Figure~\ref{fig:PlanarizatoinCrossLink}. In this example, when node $S$ needs to send data packets to destination $D$, it first forwards the packet to next closer neighbor $N$ which faces void and hence, creates a local planner graph. The obstacle shown in Figure~\ref{fig:PlanarizatoinCrossLink} hides $D$ from both $N$ and $C$ and as a result, a link between these two nodes is created which crosses the link between $D$ and $G$. Then, based on face routing and right hand rule, $N$ returns the packet back to $S$ which in turn forwards it to $B$. From $B$, the packet is forwarded to $C$ and then to $N$ where it loops. 

This means that face routing cannot always forward packets when they face void even when alternative valid paths exist. We address this problem and present an algorithm which uses a simple backtracking technique to route around voids as explained below.

When pure greedy forwarding is used, packets are dropped when they face voids making recovery procedures necessary. The existing protocols mainly depend on face routing (e.g.,GPSR)~\cite{BradKarp_Baban_2000} to recover from voids. However, in addition to failing to go around voids in some scenarios as explained above, face routing can incur high processing cost and high end-to-end delay. 

We address these issues and propose Greedy Routing Protocol with Backtracking for MANETs (GRB). GRB is a novel and simple position based routing protocol which allows each node to forward data packets to its best neighbor possible until the destination is reached. Unlike GPSR, GRB uses less computation to determine the next hop on the route.

\begin{figure*}[ht]
	 \centering
	 \includegraphics[width=0.7\textwidth]{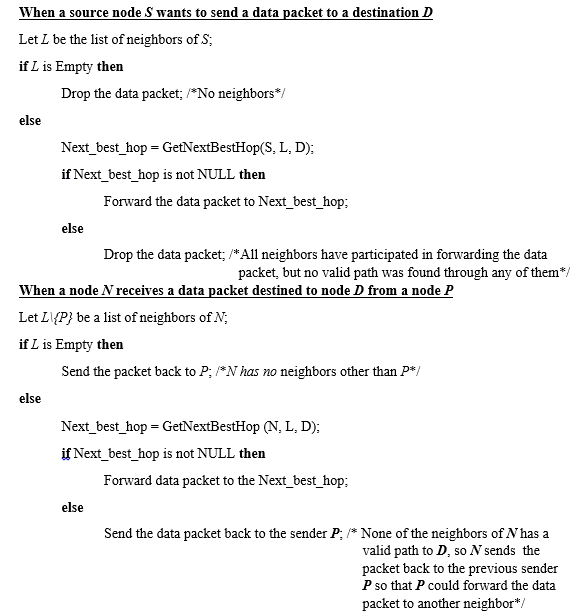}
		 \caption{GRB data forwarding (Sending/Receiving Data Packets).}
	  \label{fig:FormalDescription1}		
 \end{figure*}

\begin{figure*}[ht]
	 \centering
	 \includegraphics[width=0.7\textwidth]{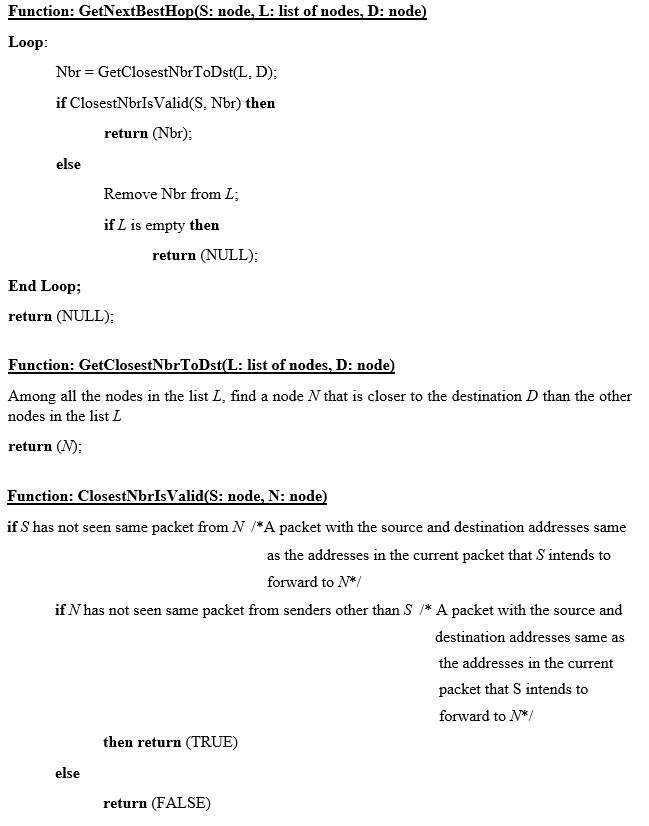}
		 \caption{GRB data forwarding (Functions).}
	  \label{fig:FormalDescription2}		
 \end{figure*}

\subsection{Basic Idea Behind Our GRB Protocol}\label{basicidea}

GRB routes data packets either in forwarding mode (greedy mode or simple
forwarding) or in backtracking mode. When a sender/intermediate node
$S$ wants to send/forward a packet to a destination $D$, it picks its $best$
$neighbor$ $N1$ and sends the packet to $N1$. The best neighbor $N1$
is determined as follows: It picks the neighbor that is closest to the
destination than any other neighbor; note that this neighbor may not be closer to the
destination than $S$ itself because $S$ may be facing a void. If the
packet backtracks from this node to $S$, it picks the one that is
closest to the destination among the remaining neighbors, and this
process continues until all neighbors have been tried; if it cannot
forward the packet through any of its neighbors, it sends the packet
back to the node from which it received the packet.  Every node on the
path uses the same strategy to forward packets. Note that if the node
picked is closer to the destination than $S$, then the forwarding is
implicitly greedy; otherwise, forwarding is around the void.
In this protocol, a
source node drops data packets if it has no neighbors, it
tried all the neighbors to forward the packet and failed, or the number
of times the packet backtracked reached a predetermined threshold.

Formal description of the protocol for data forwarding and finding
the next best hop is given in Figures~\ref{fig:FormalDescription1} and ~\ref{fig:FormalDescription2}.


\subsection{Assumptions} 

We assume that all nodes have the same transmission range (i.e., all links are bidirectional). We also assume that each node is equipped with a GPS and each node can get the location of the destination node through an available Location Service. In the following subsections we describe our protocol in detail.

\subsection{Data Structures Used in the Protocol} 
Each node maintains the following two tables.

\begin{itemize}

\item {\textbf{Neighbor Table.}}

Each node sends a $HELLO$ packet to all  its neighbors in each time
interval $T$. This $HELLO$ packet includes the node's id as well as
its position. To minimize collision of $HELLO$ packets due to
concurrent transmissions, we jitter each $HELLO$ packet transmission
interval  by $R$ milliseconds between two successive transmissions of
$HELLO$ packets so that each node  transmits $HELLO$ packets at a random time 
chosen in the interval [$T-R$, $T+R$]. When a node receives a $HELLO$
packet, it creates in its Neighbor Table an entry  containing neighbor
identifier (NbrID), neighbor position, and lifetime if that neighbor
is not in the table;  the lifetime is updated if already there
is an entry corresponding to that neighbor. If a node does not receive $HELLO$ packets for a time longer than  $2T$ from a neighbor node, it assumes the neighbor has moved and removes the associated entry from  the table.

\item {\textbf{Seen Table.}}

This table helps  picking  \textbf{\emph{best neighbor}} for
forwarding packets to the destination. For that purpose, when a node
receives a data packet, it stores the information about the packet in
its Seen Table. As shown in Table~\ref{tab:SeenN3F2}, each record of
this table contains five fields namely, neighbor ID (NbrID), source
address (Src), destination address (Dst), flag (Flag), and lifetime
(Lifetime). $NbrID$ is the address of the neighboring node that has
sent the packet, forwarded the packets, or the node from which the
packet has backtracked. $Src$ contains the address of the source
node that generated the data packet. $Flag$ indicates whether the
received packet is a new packet (i.e., forwarding mode) or it has
backtracked from a neighboring node (i.e., backtracking mode). The
flag is set to $FALSE$ when the packet is in forwarding mode and set
to $TRUE$ when it has backtracked. The lifetime field specifies the
lifetime  of the associated record in the Seen Table. When a node
receives a data packet, it creates an entry in its Seen Table if the
packet is new. However, if it has received a data packet with the same
source and destination addresses from the same neighboring node, then
it updates the lifetime of the associated record. On the other hand,
when the lifetime expires, the associated record   is removed from the table.

\end{itemize}

\subsection{Sending and Forwarding Packets}\label{SendingData}

Each node can send, forward, and/or receive data packets. When a node
has data packets to send/forward and the destination node is not one of its
neighbors, it picks the best neighbor using the method described in Sect.~\ref{basicidea} and sends/forwards
the packet to that neighbor. Since our protocol does not enforce the next-hop $N$ to be
closer to the destination than the sender $S$, $N$  is either closer
to $D$ than $S$ (i.e., Greedy mode), or farther to $D$ than
$S$. However, the next-hop $N$ must be closer to $D$ than any other
neighbor that has not seen a packet to the same source-destination
pair according to their Seen Table.  Before forwarding the packet to
$N$, the source or intermediate node $S$ does the following:

\begin{figure}[h]
	 \centering
	 \includegraphics[width=0.74\textwidth]{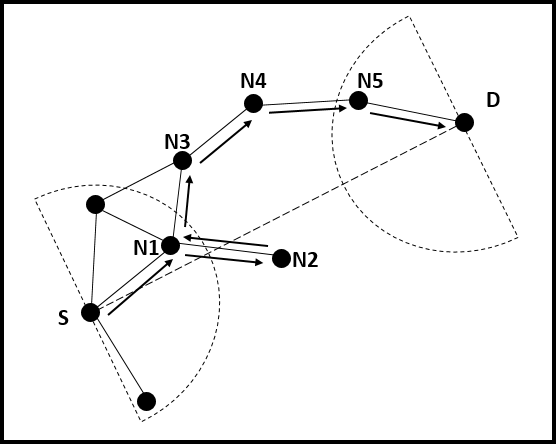}
		 \caption{Data forwarding Example.}
	  \label{fig:BackeTrackExample}		
 \end{figure}

\begin{itemize}

\item {\textbf{$S$ looks up its Seen Table for $N$.}} 

If $S$ has a record for $N$ with the same source and destination addresses as that in the packet, then it considers $N$ as invalid next hop for that packet and picks another neighboring node as the next hop. This means that $S$ has received this packet from $N$ which is either a new packet (i.e., flag is FALSE) or a backtrack packet (i.e., flag is TRUE). Therefore, it cannot forward the packet to that node because that results in a loop. For example, in Figure~\ref{fig:BackeTrackExample}, if the node $N3$ receives a data packet from node $N1$, it creates an entry in its Seen Table as shown in Table~\ref{tab:SeenN3F2}. This entry tells $N3$ that $N1$ is invalid next hop because it has received the packet from $N1$ and as a result, the Seen Table prevents loop between $N3$ and $N1$. However, the Seen Table of $N1$ does not have $N3$ as a neighbor node so it can forward data packets to $N3$.

\vspace{6pt}

 \begin{table*} [htbp]
	\centering
		\caption{Seen Table at Node $N3$ in Figure~\ref{fig:BackeTrackExample}}
		\begin{tabular} {|c|c|c|c|c|c|}  \hline  
			\textbf{NbrID} & \bfseries{Src} & \bfseries{Dst}  & \bfseries {Flag}  & \bfseries{Lifetime}
			\\  \cline {1-5}  
			  $N1$ &  $S$ & $D$ & False & $T$   	
 		\\ \cline {1-5}
		\end{tabular}

	\label{tab:SeenN3F2}
\end{table*}

\item {\textbf{$S$ verifies with $N$ if it is a valid next hop.}} 

If $N$ is not in the Seen Table of $S$, then $S$ sends $N$ a
verification packet, with same source-destination pair in the  header
as in the data packet's header, asking $N$ to check whether it has
seen data packets for the same source-destination pair from any of its other neighbors. When $N$
receives the verification packet, it checks its Seen Table for an
entry that has the same source and destination addresses as that in
the verification packet, with a Flag set to $False$, but with NbrID
different from the ID of $S$. If such an entry is found, it means that
$N$ has seen a packet for the same source-destination pair and it sends a
reply back to $S$ indicating that it is invalid next hop. However, if
such an entry is found but with Flag set to $True$, it means a neighbor $N1$
of node $N$ has sent back the data packet to $N$ after $N1$ failed
to forward the packet. In this case, maybe there are neighbors of
$N$ other than $N1$ that have not tried to forward the packet yet,
therefore $N$ is not considered as invalid next hop and as a result,
$N$ sends a reply back to $S$ indicating that it is a valid next hop
for that data packet. After receiving the reply from $N$, if $S$ finds
$N$ is a valid next hop, it forwards the packet to $N$. Otherwise, it
picks another neighbor as a new candidate for next hop and checks if
it is a valid next hop and so on.  For example, in
Figure~\ref{fig:BackeTrackExample}, when $N1$ needs to send a data
packet to $N3$, it sends a verification packet to $N3$. $N3$ checks
its Seen Table for an entry with NbrID set to any ID other than $N1$,
same Src and Dst values as those in the verification packet, and Flag
set to $False$. However, $N3$ does not have such entry in its Seen
Table (refer to Table~\ref{tab:SeenN3F2}), hence $N3$ sends a positive
reply to $N1$, and  $N1$ forwards the  packet to $N3$. If a node
finds all its neighbors are invalid next hops, then the packet is sent
back to the node from which it was received.
\item {\textbf{Packet backtracks.}} 

A packet backtracks from the current node to its sender in the following two cases:

\begin{enumerate}

 \item All the neighbors of the current (intermediate) node have seen that packet. This means none of the neighbors could forward the  packet.

 \item The current node has no neighbors other than the
 sender. For example, in Figure~\ref{fig:BackeTrackExample}, $N2$ has
 no neighbors other than $N1$ which sent the packet to it. Therefore,
 the packet backtracks  to $N1$ and $N1$ inserts a new entry to its
 Seen Table as shown in Table~\ref{tab:SeenN1F2}.  The Flag of the new
 entry (i.e., second row) is set to $True$ which means that from the
 perspective of $N1$, $N2$ is considered invalid next hop for that
 packet. Hence, when $N1$ tries to  pick next hop for the same
 destination next time, it will not pick $N2$  as long as the Lifetime
 of the associated entry (i.e., second row in Table~\ref{tab:SeenN1F2}) in the Seen Table of $N1$ has not expired.

\end{enumerate}

\begin{table*} [htbp]
	\centering
		\caption{Seen Table at Node $N1$ in Figure~\ref{fig:BackeTrackExample}}
		\begin{tabular} {|c|c|c|c|c|c|}  \hline  
			\textbf{NbrID} & \bfseries{Src} & \bfseries{Dst}  & \bfseries {Flag}  & \bfseries{Lifetime}
			\\  \cline {1-5}  
			  $S$ &  $S$ & $D$ & False & $T1$   
			\\  \cline {1-5}  
			  $N2$ &  $S$ & $D$ & True & $T2$  
 		\\ \cline {1-5}
		\end{tabular}

	\label{tab:SeenN1F2}
\end{table*}

 \begin{table*} [htbp]
	\centering
		\caption{Topology used for Simulation}
		\begin{tabular} {|c|c|c|c|}  \hline  
			\textbf{Nodes} & \bfseries{Network Area} & \bfseries{CBR Flows}  & \bfseries {Packets Sent}
			\\  \cline {1-4}  
			\{50,75,100,125,150,175,200,225,250,300\} &  1500m X 1500m & 30 & 8780   		
			\\ \cline {1-4}
			50 & 1500m X 300m & 30 & 8780  
  		\\ \cline {1-4}
		 112 & 2250m X 450m & 30 & 8780  
			\\ \cline {1-4}
		 200 & 3000m X 600m & 30 & 8780  
			\\ \cline {1-4}
		\end{tabular}

	\label{tab:Experiment1}
\end{table*}

\item{\textbf{Packet is dropped.}} A packet is dropped by a source node when all the neighbors have been identified as invalid next hops or the source node has no neighbors.

\end{itemize}

\section{Performance Analysis}\label{Analysis}
In this section, we present the performance evaluation results of GRB
compared to AODV~\cite{CalesEPS_Baban_1999}, DSR~\cite{johnsondm_Baban_1996}, and GPSR~\cite{BradKarp_Baban_2000}. We first
describe the simulation environment and then discuss the simulation
results. We simulated GRB, AODV, and DSR on a variety of network
topologies to compare the performance. We also compared the
performance of  GRB with that of GPSR using the 
results provided in  GPSR~\cite{BradKarp_Baban_2000}.

\subsection{Simulation Environment}
We used
GloMoSim~\cite{glomosim_Baban_1998}, a network-simulation tool for
studying the performance of routing protocols for ad-hoc networks,
for evaluating the performance of GRB. We chose IEEE 802.11 and IP as
the MAC  and network layer
protocols, respectively. All nodes have a
fixed transmission range of 250 m. We used the implementation of AODV and DSR
that comes with the GloMoSim 2.0.3 package to compare their performance
with GRB. We ran several simulations on two different sets of traffic
flows. The simulations run in different terrain areas are shown in
Table~\ref{tab:Experiment1}; each simulation lasted for 900 seconds of
simulated time. The nodes were  distributed uniformly at random in the network area. We used the following four metrics to evaluate performance:

\begin{enumerate}

\item Packet Delivery Ratio: Measures the success rate of delivered data packets.
\item End-to-End Delay: Average time a packet takes to reach the destination node.
\item Hop Count: The number of hops a packet traverses to reach the destination.
\item Node Density: Number of nodes in the area.
\item Network Diameter: Studying the effect of different network areas.

\end{enumerate}

In this experiment, we varied the number of nodes simulated from 50 to 300. Two sets of random traffic flows have been used in the simulation. The first set is 30 CBR (Constant Bit Rate) flows in which different senders generate data packets to be forwarded to destinations. Each CBR flow sends packets at speed of 2Kbps and uses 64-byte packets. Depending on the start time and end time of each sender in each flow, different number of packets are sent by different CBR flows. However, in each and every flow, each sender sends a packet every 0.25 second. Node mobility is set using random Way-point~\cite{johnsondm_Baban_1996} model. Under this model, each node travels from a location to a random destination at a random speed, the speed being uniformly distributed in a predefined range. After a node reaches its destination, it pauses for a predetermined amount of time and then moves to a new destination at a different randomly chosen speed. In our simulation, the speed randomly chosen lies between 0 and 20 meters/second. In order to study how mobility affects the performance of the routing protocols, we selected pause times of 0, 20, 30, 40, 60, 80, 100, and 120 seconds. When the pause time is 0 seconds, every node moves continuously. As the pause time increases, the network approaches the characteristics of a fixed network. The second set consists of 20 CBR flows which has 20 sender nodes generating packets at a speed and size same as that in the first set.

\subsection{Packet Delivery Ratio}

\begin{figure}[h]
\centering
\begin{minipage}[b]{0.45\linewidth}
\includegraphics[width=0.99\textwidth]{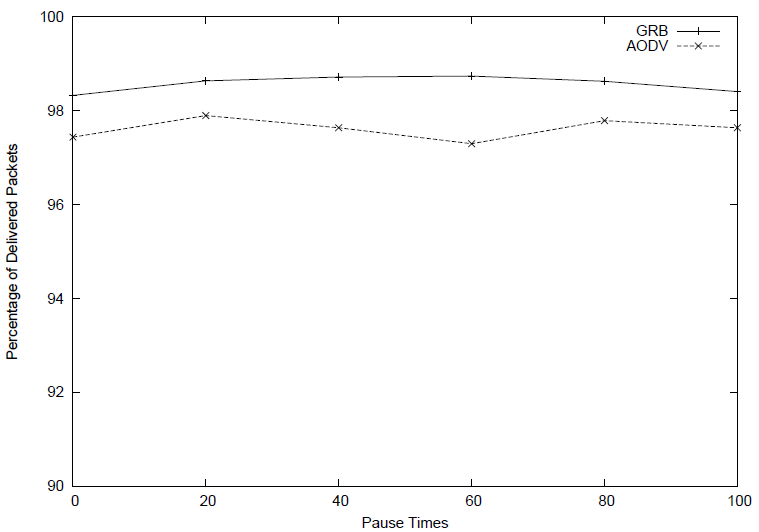}
\caption{Packet Delivery Ratio (50 Nodes, 30-CBR Flows, network area (1500m x 300m)), GRB compared with AODV. }
\label{fig:DeliveryG1}
\end{minipage}
\quad
\begin{minipage}[b]{0.45\linewidth}
\includegraphics[width=0.99\textwidth]{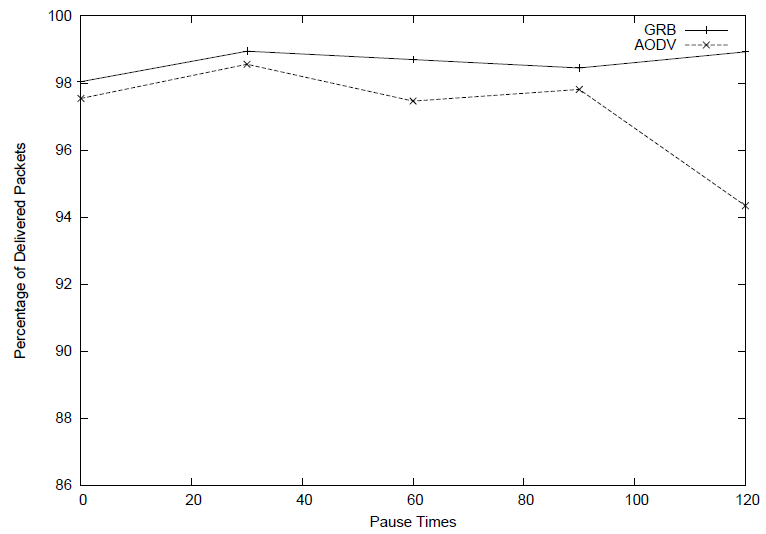}
\caption{Packet Delivery Ratio (50 Nodes, 20-CBR Flows, network area (1500m x 300m)), GRB compared with AODV.  }
\label{fig:G5Delivery}
\end{minipage}
\end{figure}

\begin{figure}[h]
\centering
\begin{minipage}[b]{0.45\linewidth}
\includegraphics[width=0.99\textwidth]{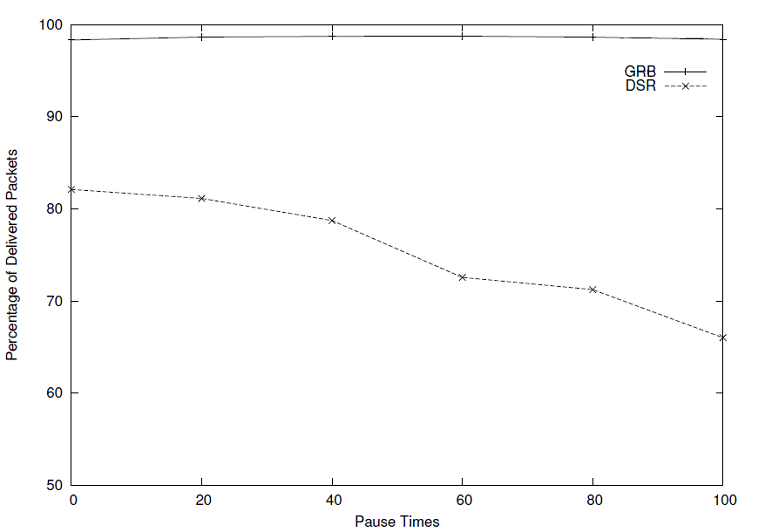}
\caption{Packet Delivery Ratio (50 Nodes, 30-CBR Flows, network area (1500m x 300m)), GRB compared with DSR. }
\label{fig:DeliveryG1DSR}
\end{minipage}
\quad
\begin{minipage}[b]{0.45\linewidth}
\includegraphics[width=0.99\textwidth]{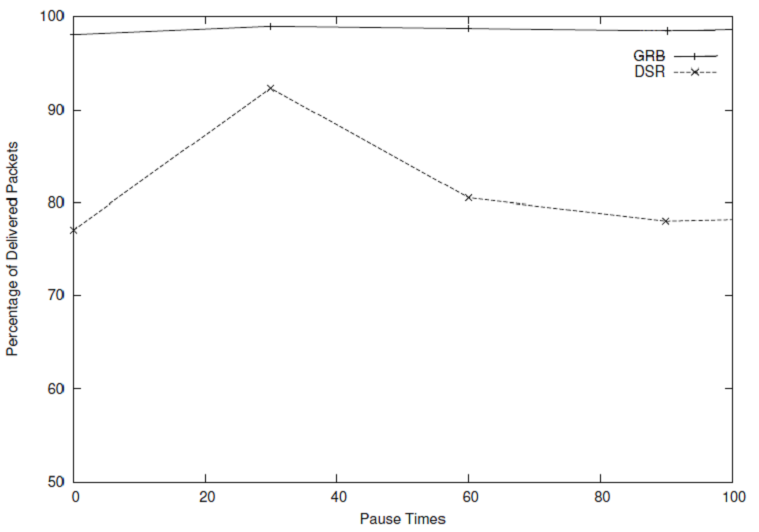}
\caption{Packet Delivery Ratio (50 Nodes, 20-CBR Flows, network area (1500m x 300m)), GRB compared with DSR.  }
\label{fig:G5DSRDelivery}
\end{minipage}
\end{figure}

The overall average packet delivery ratio for DSR, AODV, and GRB are 55.52\%, 97.38\% and 98.60\%, respectively. We select CBR flows randomly, hence it is not known whether there is a valid path between the source node and the destination node in each flow. Higher number of packets (refer to Table~\ref{tab:Experiment1}) imposes higher demand on routing protocols as higher traffic is generated between source and destination pairs.  GRB finds next hops locally with the most up to date location information of the nodes involved in the forwarding process. It simply picks next hops based on Seen Tables to forward data packets which results in few control packets. This makes GRB adapt locally to location changes, hence it tolerates mobility better than AODV and DSR. Therefore, GRB  delivers higher number of data packets than DSR and AODV for different pause times as shown in Figures~\ref{fig:DeliveryG1},~\ref{fig:G5Delivery},~\ref{fig:DeliveryG1DSR}, and~\ref{fig:G5DSRDelivery}. 


 \begin{figure}[h]
\centering
\begin{minipage}[b]{0.45\linewidth}
\includegraphics[width=0.99\textwidth]{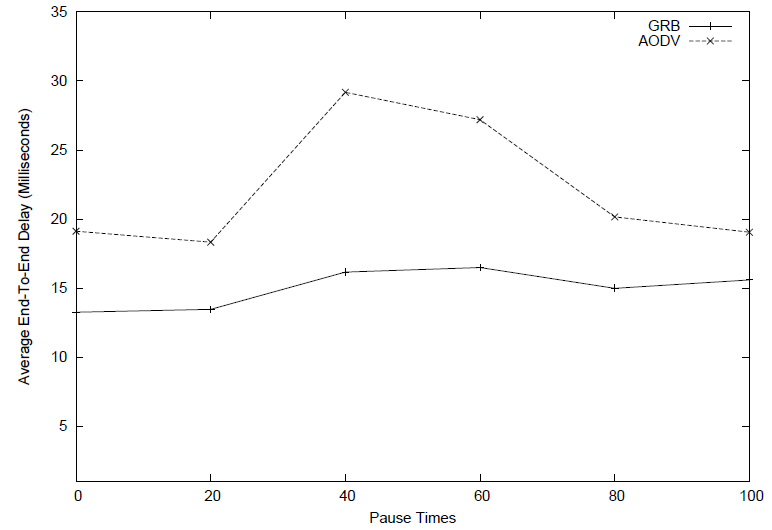}
\caption{Average End-To-End Delay (50 Nodes, 30-CBR Flows, network area (1500m x 300m)), GRB compared with AODV.}
\label{fig:DelayG2}
\end{minipage}
\quad
\begin{minipage}[b]{0.45\linewidth}
\includegraphics[width=0.99\textwidth]{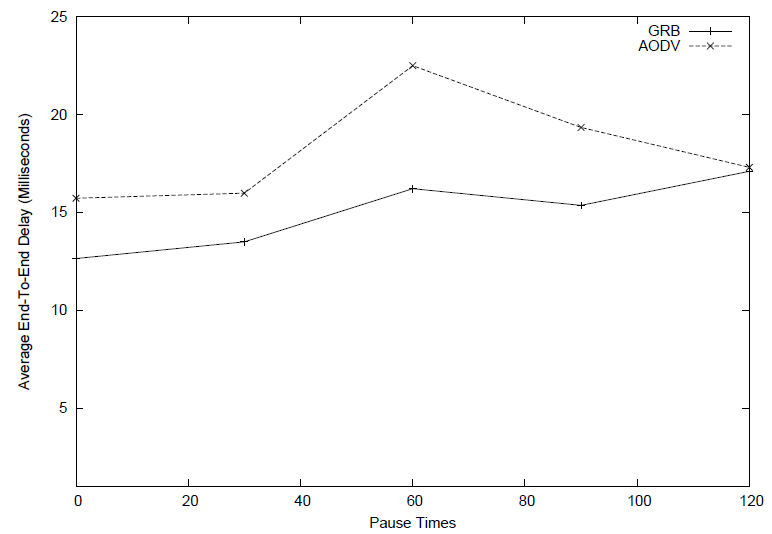}
\caption{Average End-To-End Delay (50 Nodes, 20-CBR Flows, network area (1500m x 300m)), GRB compared with AODV.}
\label{fig:G6Delay}
\end{minipage}
\end{figure}

\subsection{End-to-End Delay}

As shown in Figures~\ref{fig:DelayG2} and~\ref{fig:G6Delay}, the
overall average end-to-end delay for AODV and GRB are 22.17
milliseconds and 14.98 milliseconds, respectively. For each CBR flow,
we take the average end-to-end delay of all the packets received by
the destination node in that flow. Then, we take the average delay of
all the CBR flows. Because of its simplicity, GRB takes less time to
deliver data packets in most of the scenarios. As shown in
Figures~\ref{fig:DelayG2} and~\ref{fig:G6Delay}, packets take more
than 18 milliseconds on average to reach their destinations under
AODV; however GRB delivers packets in less than 16 milliseconds. We
can see GRB delivers packets much faster when network size and area is
moderately small (50 nodes, (1500m x 300m) area). That is because most
of the packets find greedy paths which take less calculation time and
the decision is taken quickly based on the neighbors' location
information and their status regarding whether or not they are valid
next hops. However, under AODV, data packets should wait for the route
to be set up.  Moreover, routes discovered under AODV may not be shorter than
those discovered under GRB, because GRB uses greedy approach. As a result, 
AODV  results in higher average end-to-end delay.


\subsection{Node Density}

\begin{figure}[h]
\centering
\begin{minipage}[b]{0.45\linewidth}
\includegraphics[width=0.99\textwidth]{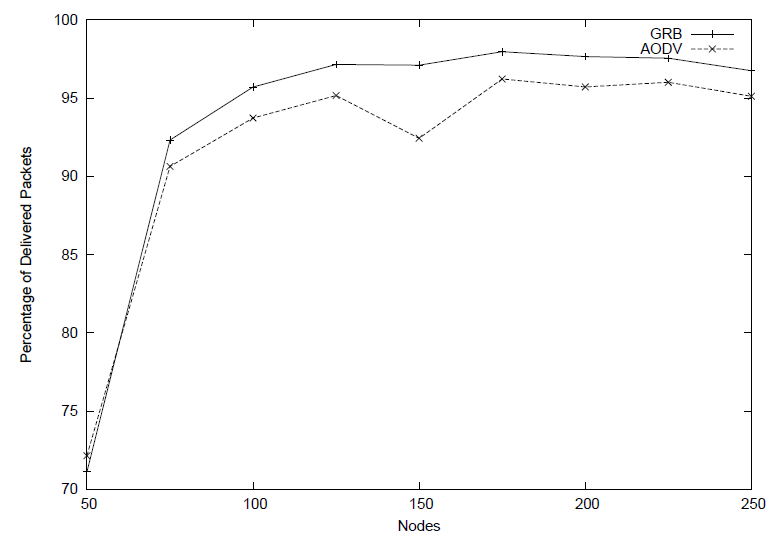}
\caption{Packet Delivery Ratio as Number of Nodes increases (Network Area (1500x1500)), GRB compared with AODV. }
\label{fig:ScalabilityDeliveryG3}
\end{minipage}
\quad
\begin{minipage}[b]{0.45\linewidth}
\includegraphics[width=0.99\textwidth]{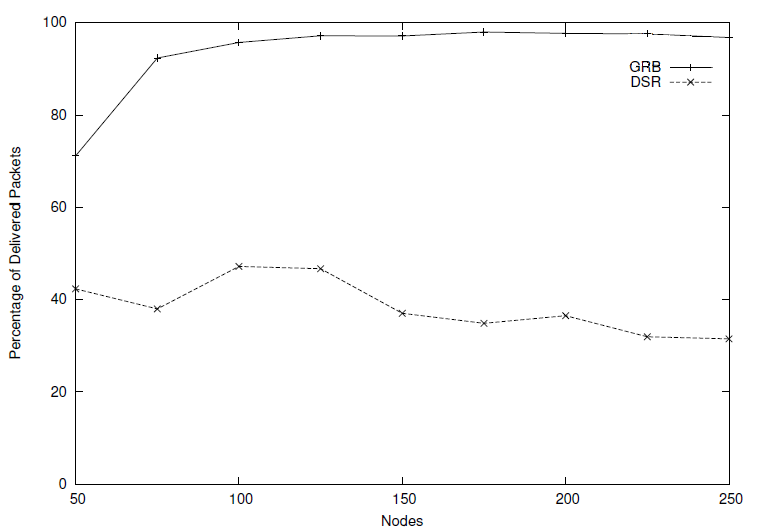}
\caption{Packet Delivery Ratio as Number of Nodes increases (Network Area (1500x1500)), GRB compared with DSR. }
\label{fig:ScalabilityDeliveryG3DSR}
\end{minipage}
\end{figure}

\begin{figure}[h]
\centering
\begin{minipage}[b]{0.45\linewidth}
\includegraphics[width=0.99\textwidth]{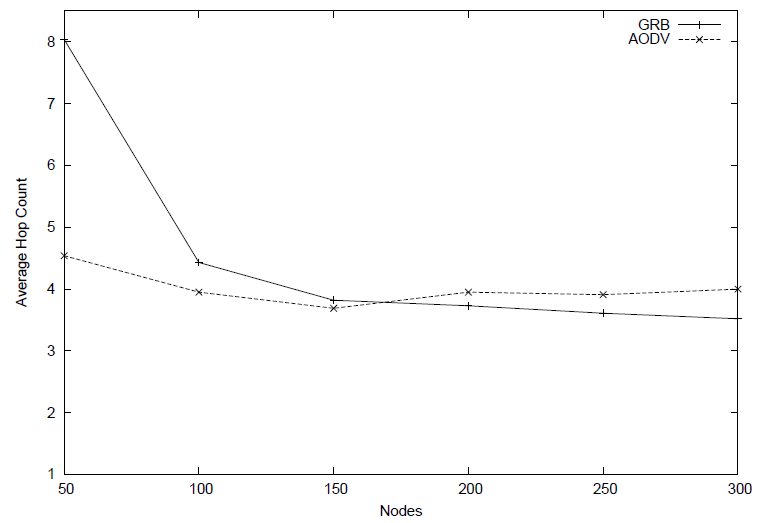}
\caption{Average Hop Count as Number of Nodes increases (Network Area (1500x1500)), GRB compared with AODV. }
\label{fig:ScalabilityHopCntG13}
\end{minipage}
\quad
\begin{minipage}[b]{0.45\linewidth}
\includegraphics[width=0.99\textwidth]{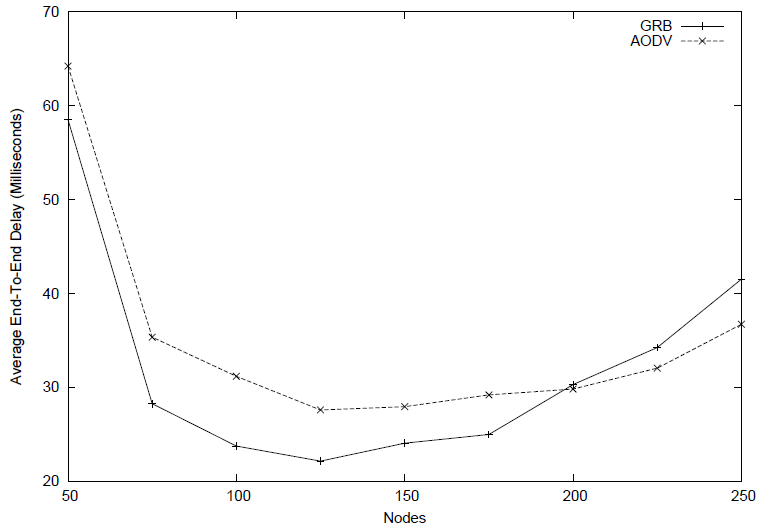}
\caption{Average End-To-End as Number of Nodes increases (Network Area (1500x1500)), GRB compared with AODV. }
\label{fig:ScalabilityDelayG4}
\end{minipage}
\end{figure}

Since our protocol uses only information about neighbors in
forwarding decision, as node density increases, GRB
keeps delivering higher fraction of data packets than AODV and DSR as shown in
Figures~\ref{fig:ScalabilityDeliveryG3} and~\ref{fig:ScalabilityDeliveryG3DSR}. That is because both AODV and DSR depend
on end-to-end route to forward data packets and that route is affected
by mobility of the nodes and the size of the network. Hence due to
mobility, more frequent link breaks occur leading to more route repair
and setup and as a result, packets are lost more frequently. However,
since the average end-to-end delay is taken only for packets that are
delivered to their destinations and because data packets follow
existing routes which decreases waiting time, AODV routes data packets
in slightly less time than GRB when number of nodes grows to more than
200 as shown in Figure~\ref{fig:ScalabilityDelayG4}. However,
GRB is faster in smaller networks (less density) because less
computation required by nodes to make forwarding decisions since nodes
have less neighbors. Average hop count is another parameter that we
measure in this simulation to show that our protocol routes data
packets with less number of hops as node density increases. For this
metric, only the successfully delivered data packets are counted in
the simulation results for both GRB and AODV. As shown in
Figure~\ref{fig:ScalabilityHopCntG13}, in smaller networks (i.e., less
than 150 nodes), AODV uses less number of hops to forward data packets
than GRB because there are more voids in sparse networks. This makes
GRB data packets to go around voids through either next best hop or
backtracking technique which makes GRB packets traverse more hops than
AODV. However, as number of nodes increases, number of voids decreases
and data packets move through greedy paths, hence GRB uses less number of hops than AODV  in dense networks. It is worth to mentioning that under GRB, average hop count is also reduced because next hop is chosen greedily.

\subsubsection{Network Diameter}
\begin{figure}[h]
\centering
\begin{minipage}[b]{0.45\linewidth}
\includegraphics[width=0.99\textwidth]{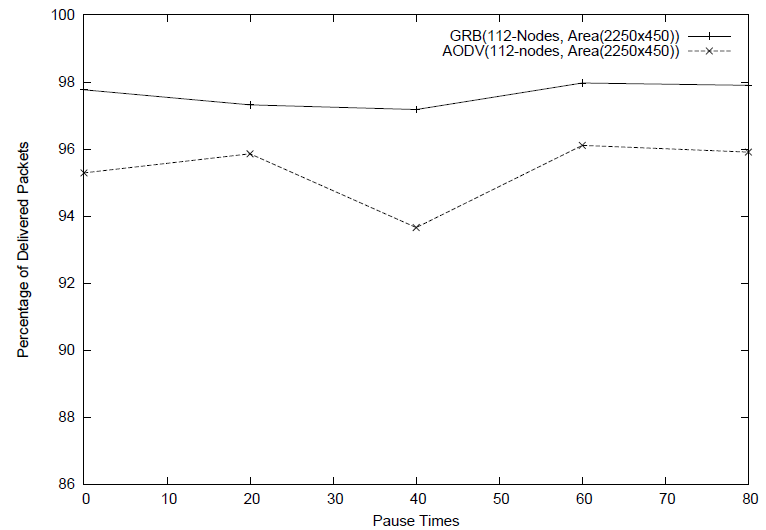}
\caption{Packet Delivery Ratio of Network Area (2250x450), 112 nodes, 30-CBR Flows, GRB compared with AODV.}
\label{fig:DiameterDeliveryG8}
\end{minipage}
\quad
\begin{minipage}[b]{0.45\linewidth}
\includegraphics[width=0.99\textwidth]{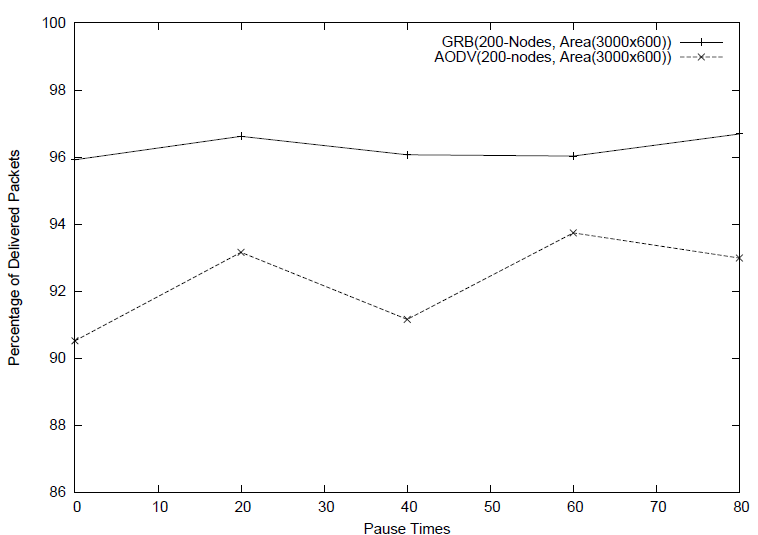}
\caption{Packet Delivery Ratio of Network Area (3000x600), 200 nodes, 30-CBR Flows, GRB compared with AODV.}
\label{fig:DiameterDeliveryG9}
\end{minipage}
\end{figure}

\begin{figure}[h]
\centering
\begin{minipage}[b]{0.45\linewidth}
\includegraphics[width=0.99\textwidth]{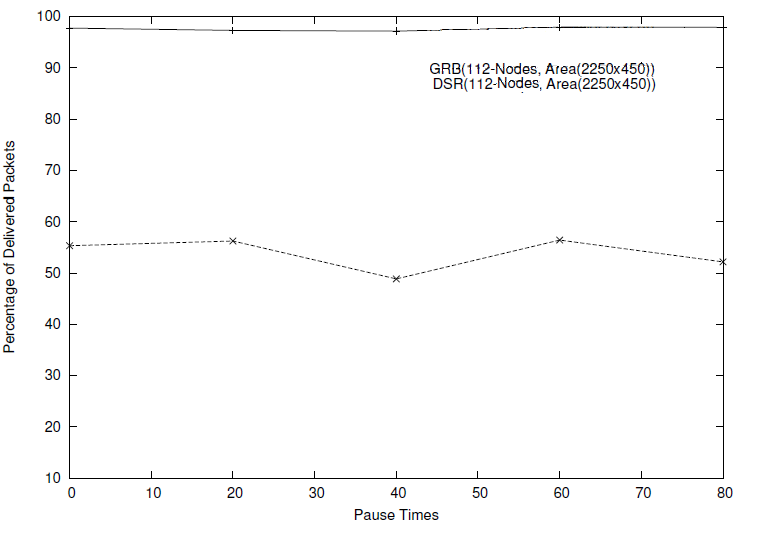}
\caption{Packet Delivery Ratio of Network Area (2250x450), 112 nodes, 30-CBR Flows, GRB compared with DSR.}
\label{fig:DiameterDeliveryG8DSR}
\end{minipage}
\quad
\begin{minipage}[b]{0.45\linewidth}
\includegraphics[width=0.99\textwidth]{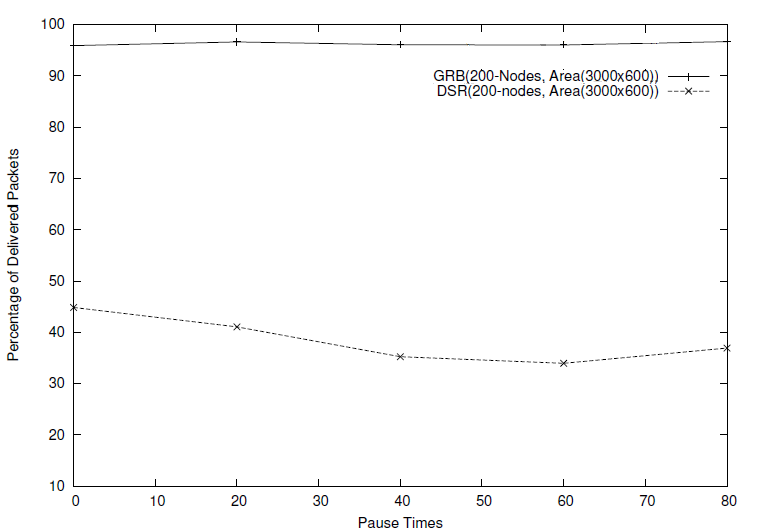}
\caption{Packet Delivery Ratio of Network Area (3000x600), 200 nodes, 30-CBR Flows, GRB compared with DSR.}
\label{fig:DiameterDeliveryG9DSR}
\end{minipage}
\end{figure}

\begin{figure}[h]
\centering
\begin{minipage}[b]{0.45\linewidth}
\includegraphics[width=0.99\textwidth]{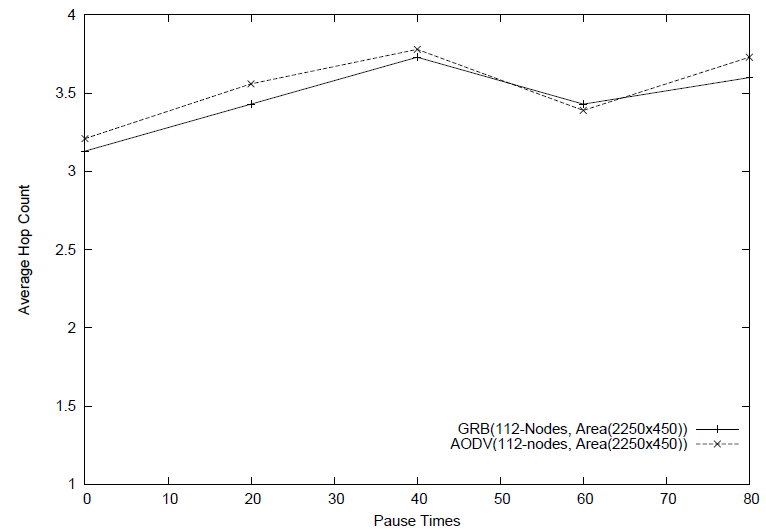}
 \caption{Average Hop Count of Network Area (2250x450), 112 nodes, 30-CBR Flows, GRB compared with AODV.}
\label{fig:DiameterHCountG11}
\end{minipage}
\quad
\begin{minipage}[b]{0.45\linewidth}
\includegraphics[width=0.99\textwidth]{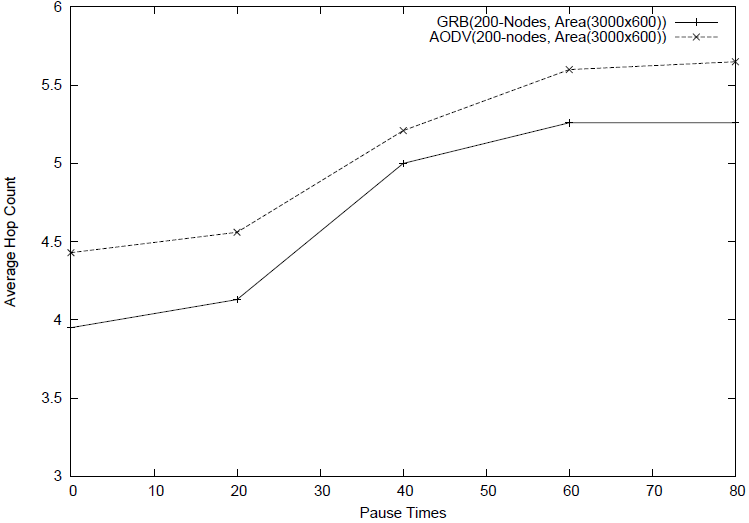}
 \caption{Average Hop Count of Network Area (3000x600), 200 nodes, 30-CBR Flows, GRB compared with AODV.}
\label{fig:DiameterHCountG12}
\end{minipage}
\end{figure}

Figures~\ref{fig:DiameterDeliveryG8} and~\ref{fig:DiameterDeliveryG9}
present packet delivery ratio results and
Figures~\ref{fig:DiameterHCountG11} and~\ref{fig:DiameterHCountG12}
present average hop count results  for  112-nodes and 200-nodes
networks with same CBR traffic and same node density for both
networks. In these simulations, the terrain area within which nodes
move are (2250x450) meters and (3000x600) meters respectively. In
these simulations, we evaluate the effect of changing network diameter
on success rate and hop count for both GRB and AODV. The probability
of a route breaking increases as the route grows longer. GRB's packet
delivery ratio is higher than AODV and DSR under all pause times on
larger networks; this is because GRB incurs no penalty when the
length of the  route increases since  GRB uses only local topology 
information to find the next best hop.  Moreover, GRB recovers from loss of
neighbor (next hop) instantaneously by simply finding another
candidate next hop which will take over the forwarding
process. However, AODV's percent delivery ratio decreases considerably
as the network diameter increases because it needs to maintain longer
end-to-end routes. DSR incurs higher traffic overhead in wider
networks since it needs to maintain longer end-to-end source routes,
hence its success rate decreases accordingly and it is much lower than
that of GRB as shown in Figures~\ref{fig:DiameterDeliveryG8DSR}
and~\ref{fig:DiameterDeliveryG9DSR}. For the hop count metric, we
calculate the average of all the received packets by all the
destination nodes under all the flows for both GRB and AODV routing
protocols. In small areas, AODV traverses less paths in higher
mobility rates (i.e., lower pause times); however, GRB uses less hop
counts when node mobility decreases (i.e., higher pause times) because
the seen table entries will be more accurate as nodes remain for
longer times in there destinations before moving to another
destination. This gives GRB more chances to direct data packets
through valid paths. In wider networks (i.e., larger diameter), GRB
uses less hop counts than AODV for all the pause times (i.e., for low
and high mobility rates) because in such networks, AODV suffers from
more route breaking occurred due to longer end-to-end paths from
source to destination nodes.

 \begin{table} [htbp]
	\centering
		\caption{Input Settings and Corresponding Results for both GRB and GPSR}
		\begin{tabular} {|c|c|c|c|c|}  \hline  
			\textbf{Nodes} & \bfseries{Network Area} & \bfseries{PT(s)}  & \bfseries {SPDR(GRB)}& \bfseries {SPDR(GPSR)}
			\\  \cline {1-5}  
			50 & 1500m X 300m & 0 & 98.98 & 97.04   		
			\\ \cline {1-5}
			50 & 1500m X 300m & 60 & 99.07 & 98.16  
  		\\ \cline {1-5}
		 112 & 2250m X 450m & 0 & 98.00 & 97.50  
			\\ \cline {1-5}
		 112 & 2250m X 450m & 60 & 98.54 & 98.25  
			\\ \cline {1-5}
		 200 & 3000m X 600m & 0 & 97.02 & 95.00  
			\\ \cline {1-5}
		 200 & 3000m X 600m & 60 & 96.84 & 97.50  
		 \\ \cline {1-5}
		\end{tabular}

	\label{tab:Settings_GRB}
\end{table}

 \begin{figure}[h]
	 \centering
	 \includegraphics[width=0.6\textwidth]{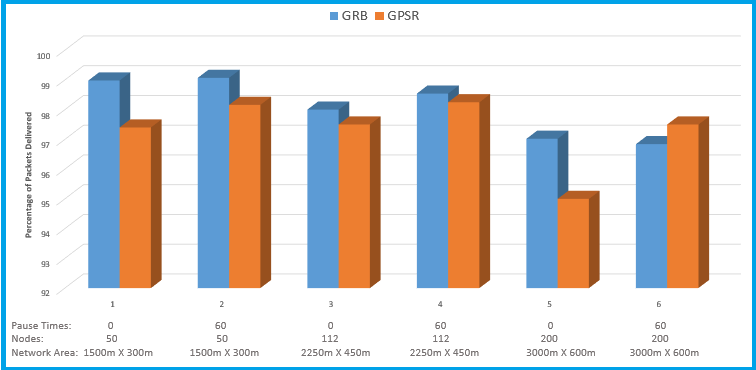}
	 \caption{ GRB Vs. GPSR.}
	  \label{fig:GRBvsGPSR}
 \end{figure}

\subsection{GRB Vs. GPSR}

Even though we didn't simulate GPSR, we used the performance
results published for GPSR in~\cite{BradKarp_Baban_2000} and the results we obtained
for GRB to compare the performance of the two protocols. As stated
in~\cite{BradKarp_Baban_2000}, GPSR performance evaluation counts only those packets for which a path
exists to the destination. We used the same input settings as those
used for GPSR to compare the success rate. The settings are: 50 nodes,
30 CBR flows, pause times (PT) (0, 30, 60, and 120) seconds, area
(1500x300) meters, and node density ($1 node/9000 m^2$). From the
results presented in GPSR~\cite{BradKarp_Baban_2000}, successful packet delivery
rate (SPDR) of GPSR ranges from 95\% to 99.10\% for pause time 0 second,
while GRB successfully delivers 98.93\% of the total packets
sent. When pause time is 30 seconds, GPSR achieves success rates
between 98.20\% to 99.70\% whereas GRB achieves a success rate of 99.33\%. When pause
time is 60 seconds, GPSR delivers from 98.70\% to 99.40\% of data
packets sent, while GRB delivers 99.04\% of all packets. Finally, when
pause time is 120 seconds, GPSR's packet delivery rate ranges from
98.60\% to 99.40\% and GRB's packet delivery  rate is 99.28\%. The result of the comparison is shown in Table~\ref{tab:Settings_GRB}.

\begin{figure}[h]
	 \centering
	 \includegraphics[width=0.60\textwidth]{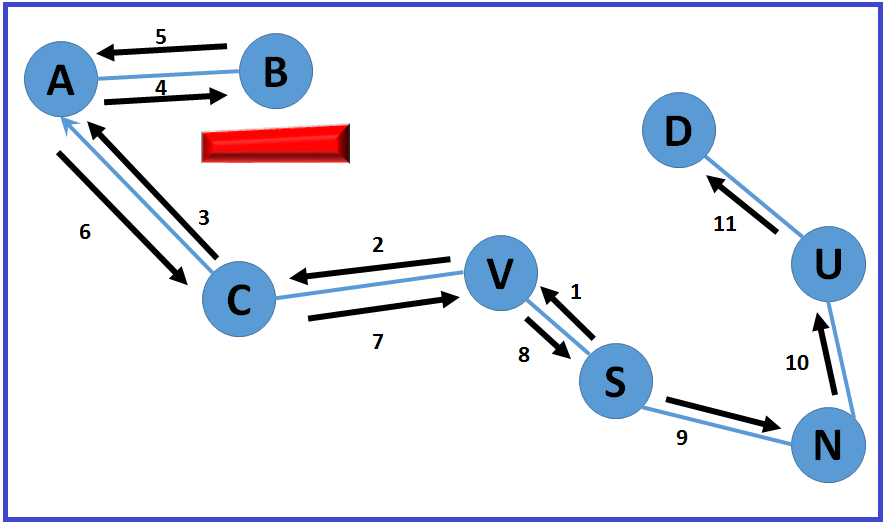}
	 \caption{GRB Succeeds when Unidirectional Links Cause Routing Failure.}
	  \label{fig:GRBsucceedsUnidirectional}
 \end{figure}

\begin{figure}[h]
	 \centering
	 \includegraphics[width=0.62\textwidth]{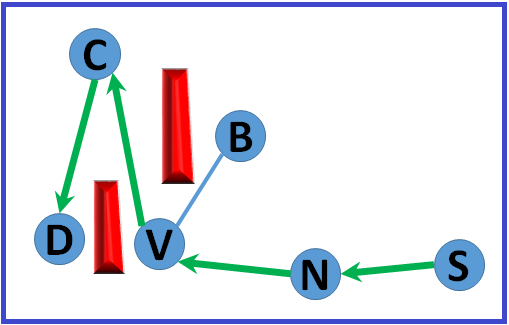}
	 \caption{ GRB Succeeds when Disconnected Links Cause Routing Failure.}
	  \label{fig:GRBsucceedsUDisconnected}
 \end{figure}

\begin{figure}[h]
	 \centering
	 \includegraphics[width=0.50\textwidth]{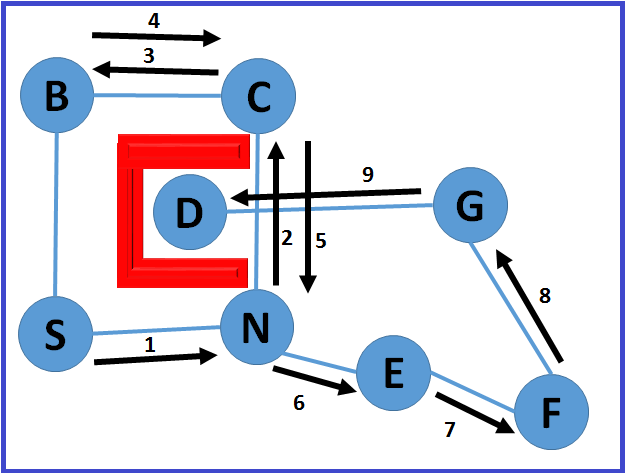}
	 \caption{ GRB Succeeds when Cross Links Cause Routing Failure.}
	  \label{fig:GRBsucceedsCrossLinks}
 \end{figure}

As shown in Figure~\ref{fig:GRBvsGPSR}, for 50 and 112 nodes, pause times 0 and 60 seconds, areas (1500x300) and (2250x450) meters, we see that GRB performs better than GPSR with respect to Successful Packet Delivery Ratio (SPDR).  
We noticed that the performance of our protocol in successfully
delivering data packets is almost same as that of
GPSR in some cases and higher than GPSR in other cases. However, when greedy routing fails due to a void in the
direction of the destination,  GPSR  has to planarize the
local network graph and use it to route around voids. Planarizing the
graph results in  computation overhead as well as routing failure. Figures~\ref{fig:GRBsucceedsUnidirectional},~\ref{fig:GRBsucceedsUDisconnected}, and~\ref{fig:GRBsucceedsCrossLinks} show how GRB successfully routes around voids where GPSR could not because of the planarization and face routing problems discussed in Section~\ref{problemstatement}. When face routing fails because of unidirectional links as shown in Figure~\ref{fig:PlanarizatoinUnidirectional}, under GRB, the first packet follows the links labeled (1,2,3,4,5,6,7,8,9,10,11) to reach the destination as shown in Figure~\ref{fig:GRBsucceedsUnidirectional}. However, the remaining packets will be routed via the links labeled (9,10,11) because the first packet backtracked from $V$ to $S$. Hence, in the \textit{Seen Table} of the source $S$, it is stated that node $V$ is an invalid next hop for the destination $D$.

When planarization of the local network graph results in disconnected links as shown in Figure~\ref{fig:PlanarizatoinDisconnected}, data packets under GRB follow the sold arrows (i.e. nodes \textit{(N,V,C)}) to reach the destination $D$ as shown in Figure~\ref{fig:GRBsucceedsUDisconnected}. When cross links caused data packets to loop as shown in Figure~\ref{fig:PlanarizatoinCrossLink}, the first packet follows the links labeled (1,2,3,4,5,6,7,8,9) under GRB as shown in Figure~\ref{fig:GRBsucceedsCrossLinks}. Moreover, the remaining packets follow the path labeled (6,7,8,9) towards $D$.

GRB neither requires planarizarion of local network graph nor it switches from greedy mode to an alternative mode when
a packet faces void; instead a node simply selects the \textit{best next hop} without imposing the condition that the next hop be closer to the destination than itself. GRB depends on the \textit{Seen Table} to determine the next hop. Therefore, GRB needs less control information which makes it better and robust. So, we can see the difference between GRB and GPSR in delivering data packets is around 0.0066\% in one scenario keeping in mind that GRB outperforms GPSR in some other scenarios. When  compared to the simplicity of GRB and the less  control overhead  GRB needs, this difference is negligible. It is worthy to recall that we select CBR flows randomly without knowing whether there exists a route corresponding to each flow. However, under GPSR, according to the authors: "Only packets for which a path exists to the destination are included in the graph".

\section{Discussion}\label{Discussion}
We note from the detailed description of the protocol
Sect.~\ref{algorithm} that data packets are forwarded greedily. When
greedy forwarding fails, GRB  picks the
next best hop based on simple heuristics without incurring large
computation overhead, unlike GPSR. A packet can come back (backtrack) to its sender/forwarder
if the next hop picked for forwarding packet could not use any of
its neighbors to forward the packet. If a packet backtracks from a
neighbor selected as next hop, that next hop is recorded in the Seen
Table; this prevents a node from selecting the same node as next hop when it has
to forward succeeding packets to the same destination. Benefits of the Seen Table: (i) It 
prevents loops; (ii) after the first packet establishes the route to
the destination, all the packets for the same source-destination pair 
follow the same route; (iii) it helps in  decreasing  hop count and  latency.

\section{Conclusion}\label{Conclusion}

In this paper, we presented GRB, a simple low-overhead position-based
routing protocol which  consistently and successfully delivers high
percentage of data packets. We compared the performance of GRB with
well known position-based protocol GPSR, the on-demand routing
protocol AODV, and the Dynamic Source Routing (DSR) protocol. Our performance evaluation shows that GRB performs as
good as GPSR (with low overhead)
and better than AODV and DSR under most scenarios. Unlike GPSR, GRB does not
need to construct planar graphs to route around voids; it simply picks the best next hop to forward the data packets, hence GRB is simple. On the other hand, it achieves comparable packet delivery ratio to GPSR and AODV, hence it is robust.

\bibliographystyle{IEEEtran}

\bibliography{MyReferences}

\begin{thebibliography}{10}
\providecommand{\url}[1]{#1}
\csname url@samestyle\endcsname
\providecommand{\newblock}{\relax}
\providecommand{\bibinfo}[2]{#2}
\providecommand{\BIBentrySTDinterwordspacing}{\spaceskip=0pt\relax}
\providecommand{\BIBentryALTinterwordstretchfactor}{4}
\providecommand{\BIBentryALTinterwordspacing}{\spaceskip=\fontdimen2\font plus
\BIBentryALTinterwordstretchfactor\fontdimen3\font minus
  \fontdimen4\font\relax}
\providecommand{\BIBforeignlanguage}[2]{{%
\expandafter\ifx\csname l@#1\endcsname\relax
\typeout{** WARNING: IEEEtran.bst: No hyphenation pattern has been}%
\typeout{** loaded for the language `#1'. Using the pattern for}%
\typeout{** the default language instead.}%
\else
\language=\csname l@#1\endcsname
\fi
#2}}
\providecommand{\BIBdecl}{\relax}
\BIBdecl

\bibitem{HuaizhiLi_Baban_2005}
H.~Li and M.~Singhal, ``{An anchor-based routing protocol with cell id
  management system for ad hoc networks},'' in \emph{Proceedings of
  International Conference on Computer Communications and Networks}, Oct 2005.

\bibitem{HShen_Baban_2013}
H.~Shen and L.~Zhao, ``Alert: An anonymous location-based efficient routing
  protocol in manets,'' \emph{IEEE Transactions on Mobile Computing}, vol.~12,
  no.~6, pp. 1079--1093, June 2013.

\bibitem{NazariTV_Baban_2013}
V.~N. Talooki, H.~Marques, and J.~Rodriguez, ``Energy efficient dynamic manet
  on-demand (e2dymo) routing protocol,'' in \emph{Proceedings of International
  Symposium and Workshops on a World of Wireless, Mobile and Multimedia
  Networks (WoWMoM)}, June 2013.

\bibitem{KKavithal_Baban_2013}
K.~Kavitha, K.~Selvakumar, T.~Nithya, and S.~Sathyabama, ``Zone based multicast
  routing protocol for mobile ad hoc network,'' in \emph{Proceedings of
  International Conference on Emerging Trends in VLSI, Embedded System, Nano
  Electronics and Telecommunication System (ICEVENT)}, Jan 2013.

\bibitem{JainSS_Baban_2013}
S.~Jain, A.~Shastri, and B.~K. Chaurasia, ``Analysis and feasibility of
  reactive routing protocols with malicious nodes in manets,'' in
  \emph{Proceedings of International Conference on Communication Systems and
  Network Technologies (CSNT)}, April 2013.

\bibitem{junlongLin_Baban_2006}
J.~Lin and G.-S. Kuo, ``A novel location-fault-tolerant geographic routing
  scheme for wireless ad hoc networks,'' in \emph{Proceedings of 63rd Vehicular
  Technology Conference}, May 2006.

\bibitem{RolandF_Baban_2008}
R.~Flury and R.~Wattenhofer, ``Randomized 3d geographic routing,'' in
  \emph{Proceedings of 27th IEEE international Conference on Computer
  Communications (INFOCOM)}, April 2008.

\bibitem{BradKarp_Baban_2000}
B.~Karp and H.~T. Kung, ``Gpsr: Greedy perimeter stateless routing for wireless
  networks,'' in \emph{Proceedings of the 6th Annual International Conference
  on Mobile Computing and Networking}, 2000.

\bibitem{ChiaCH_Baban_2009}
C.~C. Hsu and C.~L. Lei, ``A geographic scheme with location update for ad hoc
  routing,'' in \emph{Proceedings of Fourth International Conference on Systems
  and Networks Communications, ICSNC}, Sept 2009.

\bibitem{VMukesh_Baban_2007}
V.~C. Giruka and M.~Singhal, ``A self-healing on-demand geographic path routing
  protocol for mobile ad-hoc networks,'' \emph{Ad Hoc Netw.}, vol.~5, no.~7,
  pp. 1113--1128, Sep. 2007.

\bibitem{CalresEP_Baban_1994}
C.~E. Perkins and P.~Bhagwat, ``Highly dynamic destination-sequenced
  distance-vector routing (dsdv) for mobile computers,'' \emph{SIGCOMM Comput.
  Commun. Rev.}, vol.~24, no.~4, pp. 234--244, Oct. 1994.

\bibitem{StefanoBI_Baban_1998}
S.~Basagni, I.~Chlamtac, V.~R. Syrotiuk, and B.~A. Woodward, ``A distance
  routing effect algorithm for mobility (dream),'' in \emph{Proceedings of the
  4th Annual ACM/IEEE International Conference on Mobile Computing and
  Networking (MobiCom)}, 1998.

\bibitem{zhaoyb_Baban_2007}
Y.~Zhao, Y.~Chen, B.~Li, and Q.~Zhang, ``Hop id: A virtual coordinate based
  routing for sparse mobile ad hoc networks,'' \emph{IEEE Transactions on
  Mobile Computing}, vol.~6, no.~9, pp. 1075--1089, 2007.

\bibitem{lemmoncls_Baban_2009}
C.~Lemmon, S.~M. Lui, and I.~Lee, ``Geographic forwarding and routing for
  ad-hoc wireless network: A survey,'' in \emph{Proceedings of Fifth
  International Joint Conference on INC, IMS and IDC}.\hskip 1em plus 0.5em
  minus 0.4em\relax IEEE, 2009.

\bibitem{shobanaks_Baban2013}
M.~Shobana and S.~Karthik, ``A performance analysis and comparison of various
  routing protocols in manet,'' in \emph{Proceedings of International
  Conference on Pattern Recognition, Informatics and Mobile Engineering
  (PRIME)}.\hskip 1em plus 0.5em minus 0.4em\relax IEEE, 2013.

\bibitem{cadgerfc_Baban_2013}
F.~Cadger, K.~Curran, J.~Santos, and S.~Moffett, ``A survey of geographical
  routing in wireless ad-hoc networks,'' \emph{IEEE Communications Surveys \&
  Tutorials}, vol.~15, no.~2, pp. 621--653, 2013.

\bibitem{Kimyrkb_Baban_2005}
Y.-J. Kim, R.~Govindan, B.~Karp, and S.~Shenker, ``On the pitfalls of
  geographic face routing,'' in \emph{Proceedings of the Joint Workshop on
  Foundations of Mobile Computing}, ser. DIALM-POMC '05.\hskip 1em plus 0.5em
  minus 0.4em\relax New York, NY, USA: ACM, 2005.

\bibitem{freyFace_Baban_2006}
H.~Frey and I.~Stojmenovic, ``On delivery guarantees of face and combined
  greedy-face routing in ad hoc and sensor networks,'' in \emph{Proceedings of
  the 12th annual international conference on Mobile computing and
  networking}.\hskip 1em plus 0.5em minus 0.4em\relax ACM, 2006.

\bibitem{zhoublyzg_Baban_2008}
B.~Zhou, Y.-Z. Lee, and M.~Gerla, ``Direction assisted geographic routing for
  mobile ad hoc networks,'' in \emph{Proceedings of MILCOM Military
  Communications Conference}.\hskip 1em plus 0.5em minus 0.4em\relax IEEE,
  2008.

\bibitem{gerlamlyzz_Baban_2006}
M.~Gerla, Y.-Z. Lee, B.~Zhou, J.~Chen, and A.~Caruso, ``Direction forwarding
  for highly mobile, large scale ad hoc networks,'' in \emph{Challenges in Ad
  Hoc Networking}.\hskip 1em plus 0.5em minus 0.4em\relax Springer, 2006, pp.
  357--366.

\bibitem{GuangyuGr_Baban_2000}
G.~Pei, M.~Gerla, and T.-W. Chen, ``Fisheye state routing: a routing scheme for
  ad hoc wireless networks,'' in \emph{Proceedings of IEEE International
  Conference on Communications}, 2000.

\bibitem{XLINMN_Baban_2012}
X.~Li, N.~Mitton, A.~Nayak, and I.~Stojmenovic, ``Localized load-aware
  geographic routing in wireless ad hoc networks,'' in \emph{Proceedings of
  IEEE International Conference on Communications (ICC)}, June 2012.

\bibitem{StojmenovicI_Baban_2006}
I.~Stojmenovic, ``Localized network layer protocols in wireless sensor networks
  based on optimizing cost over progress ratio,'' \emph{IEEE Network}, vol.~20,
  no.~1, pp. 21--27, Jan 2006.

\bibitem{macintoshsfl_Baban2012}
A.~Macintosh, M.~Ghavami, M.~F. Siyau, and S.~Ling, ``Local area network
  dynamic (landy) routing protocol: A position based routing protocol for
  manet,'' in \emph{Proceedings of 18th European Wireless Conference}.\hskip
  1em plus 0.5em minus 0.4em\relax VDE, 2012.

\bibitem{CalesEPS_Baban_1999}
C.~E. Perkins and E.~M. Royer, ``Ad-hoc on-demand distance vector routing,'' in
  \emph{Proceedings of Second IEEE Workshop on Mobile Computing Systems and
  Applications, WMCSA}, Feb 1999.

\bibitem{johnsondm_Baban_1996}
D.~B. Johnson and D.~A. Maltz, ``Dynamic source routing in ad hoc wireless
  networks,'' in \emph{Mobile computing}.\hskip 1em plus 0.5em minus
  0.4em\relax Springer, 1996, pp. 153--181.

\bibitem{gabrielks_Baban1969}
K.~R. Gabriel and R.~R. Sokal, ``A new statistical approach to geographic
  variation analysis,'' \emph{Systematic Biology}, vol.~18, no.~3, pp.
  259--278, 1969.

\bibitem{toussaintGo_Baban_1980}
G.~T. Toussaint, ``The relative neighbourhood graph of a finite planar set,''
  \emph{Pattern recognition}, vol.~12, no.~4, pp. 261--268, 1980.

\bibitem{glomosim_Baban_1998}
\BIBentryALTinterwordspacing
X.~Zeng, R.~Bagrodia, and M.~Gerla, ``Glomosim: A library for parallel
  simulation of large-scale wireless networks,'' \emph{SIGSIM Simul. Dig.},
  vol.~28, no.~1, pp. 154--161, Jul. 1998. [Online]. Available:
  \url{http://doi.acm.org/10.1145/278009.278027}
\BIBentrySTDinterwordspacing

\end{thebibliography}

\end{document}